\documentclass{article}

\pdfoutput=1

\usepackage[latin1]{inputenc}
\usepackage{amsmath}
\usepackage{amsfonts}
\usepackage{amssymb}
\usepackage{graphicx}
\usepackage{epstopdf}
\usepackage{color}
\usepackage{appendix}
\usepackage{csvsimple}
\usepackage{longtable}
\usepackage{chngpage}
\usepackage{subcaption}
\usepackage[export]{adjustbox}[2011/08/13]
\usepackage{mathrsfs}
\usepackage{a4wide}
\usepackage{psfrag}
\usepackage{multirow}
\usepackage{hhline}
\usepackage{grffile}

\newcommand{\lsim}{\mbox{\raisebox{-.9ex}{~$\stackrel{\mbox{$<$}}{\sim}$~}}}
\newcommand{\gsim}{\mbox{\raisebox{-.9ex}{~$\stackrel{\mbox{$>$}}{\sim}$~}}}
\newcommand{\calp}{\mbox{${\cal P}$}}

\begin{document}

\begin{center}
{\Large\bf An analytic treatment of Quartic Hilltop Inflation}

\bigskip

{\large Konstantinos Dimopoulos}\footnote{\tt k.dimopoulos1@lancaster.ac.uk}

\medskip

{\sl Consortium for Fundamental Physics}

{\sl Physics Department, Lancaster University}

{\sl Lancaster, LA1 4YB, UK}

\end{center}

\begin{abstract}
  Quartic hilltop inflation remains one of the most successful inflationary
  models. Yet, the expectations of early treatments of hilltop
  inflation would contradict the observations and render the model excluded.
  However, recent numerical treatment has demonstrated that quartic hilltop
  inflation actually fares well with observations. In this work, a fully
  analytic treatment of the model aims to dispel the mystery surrounding the
  behaviour of quartic hilltop inflation. The results obtained are in excellent
  agreement with numerical works on the subject, yet offer simple analytic
  formulas to calculate observables and easily test thereby quartic hilltop
  inflation, hopefully revealing information on the theoretical background.
\end{abstract}

\bigskip

The precision of cosmological observations in the last few years has grown so
high that the paradigm of inflation model-building is changed. Out are the
simple monomial chaotic models while centre-stage is reserved for plateau
models. Yet, in contrast to the large-field plateau models, hilltop inflation,
a small field family of models, seems to be doing well. However, based only on
the early expectations we would expect the model to be another failure.
What allows hilltop inflation to escape the fate of monomial chaotic models? 

The term hilltop inflation was coined by Boubekeur and Lyth in 2005
\cite{hilltop} but the model was known well before, as far back as new inflation
\cite{new}. The potential density of the model is 
\begin{equation}
V(\phi)=V_0\left[1-\lambda\left(\frac{\phi}{m_P}\right)^q\right]+\cdots\,,
\label{V0}
\end{equation}
where $\phi$ is the inflaton field,
$V_0$ is a constant density scale, $\lambda$ some parameter, $q>0$ is
typically an integer, the ellipsis denotes higher order terms which stabilise
the potential and \mbox{$m_P=(8\pi G)^{-1/2}$} is the reduced Planck mass with
$G$ being Newton's gravitational constant.
Early analytic treatment of the model suggested that the
spectral index of the curvature perturbation is \cite{lythriotto}
\begin{equation}
n_s=1-2\left(\frac{q-1}{q-2}\right)\frac{1}{N}\,,
\label{nsearly}
\end{equation}
where $N$ is the remaining e-folds of inflation when the cosmological scales
exit the horizon.

Arguably, one of the most motivated members of the family of hilltop inflation
models is quartic hilltop inflation, where \mbox{$q=4$}. Thus, in the case of
quartic hilltop inflation we would expect
\begin{equation}
n_s=1-\frac{3}{N}\,,
\label{nsnaive}
\end{equation}
Taking \mbox{$N=60$}, the above gives \mbox{$n_s=0.95$}, which is much too low
to be acceptable. The number of e-folds can decrease if there is a period of
effective matter domination after inflation, but this makes the spectrum even
redder. One can obtain the desired value \mbox{$n_s\gsim 0.96$} only when
\mbox{$N\gsim 75$}, which is way too large.\footnote{Even with a subsequent
  period of kination, lasting until the electroweak scale, $N$ cannot grow
  larger than 67.} Thus, we are temped to conclude that quartic hilltop
inflation is excluded. 

\begin{figure} [t]
\vspace{-15cm}

\centering
\includegraphics[width=17cm]{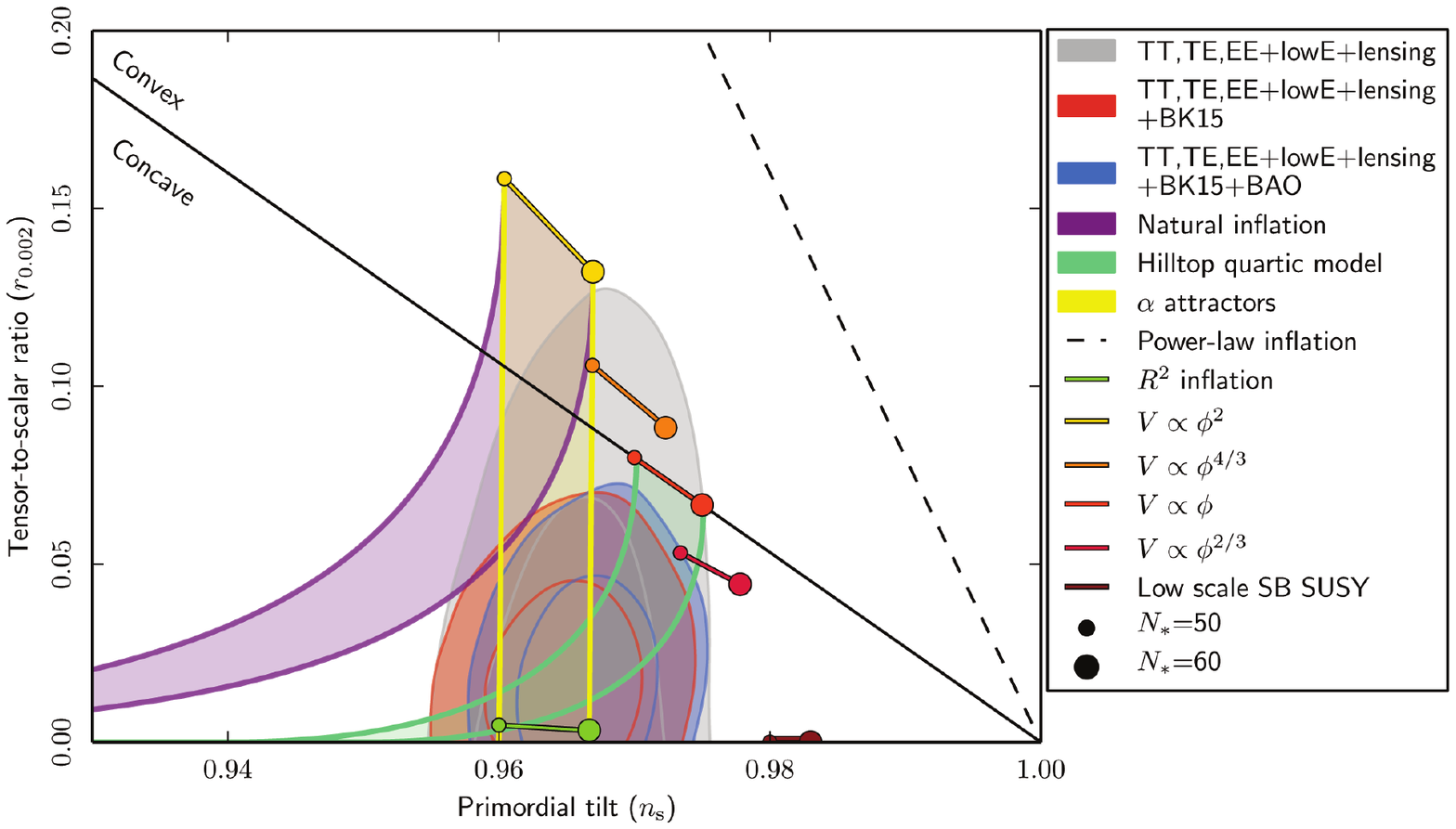} 
\caption{Diagram in the $n_s-r$ plane featuring the Planck 2018 data contrasted
  with the predictions of some prominent inflation models. The predictions of
  quartic hilltop inflation are shown with cyan convex curves (see legend). It
  can be seen that the model is successful as there is significant overlap with
  the allowed parameter space. Figure taken from Ref.~\cite{Planck2018}.}  
\label{Planck2018}
\end{figure}

However, in recent years in was gradually realised that actually hilltop
inflation fares well with observations. In fact, quartic hilltop inflation is
one of the models selected by the Planck Collaboration to feature in their
main graph (see Fig.~\ref{Planck2018}). The model has been carefully
looked into in {\sl Encyclop\ae dia Inflationaris} \cite{inflationaris},
whose results are being used by the Planck Collaboration. However, the treatment
was largely numerical and one might say obscure. It was not clear why the model
departs from the early expectations described above. A more recent work
\cite{linde} has made an effort to peer into the physical interpretation of the
behaviour of hilltop inflation, found in Ref.~\cite{inflationaris}. However,
this too is a numerical investigation.

It can be argued that numerical investigation, while it can reveal unexpected
behaviour, which was previously not realised,
is in itself rather
opaque, the computer being akin to a black box. In contrast, the clarity of
analytic calculation is unparalleled. We already know {\sl what} should come out
now, thanks to the numerical treatment. In this work, we look into the {\sl why}
quartic inflation behaves as~such.

The result in Eq.~(\ref{nsnaive}) (and Eq.~(\ref{nsearly})) is based on two
crucial assumptions. One is that the expectation value of the inflaton field
$N$ e-folds before the end of inflation is much smaller than its vacuum
expectation value (VEV), i.e.
\mbox{$\phi(N)\ll\langle\phi\rangle\sim m_P/\lambda^{1/q}$} (with \mbox{$q=4$}),
such that during inflation \mbox{$V(\phi)\simeq V_0$}.
The other assumption is that the contribution of the value of the inflaton
$\phi_{\rm end}$ when inflation ends in negligible and disregarded in the
calculation of $N(\phi)$ (which is the inverse of $\phi(N)$). It turns out that
both these assumptions are unwarranted. Below we analytically calculate
the spectral index $n_s$ and the tensor to scalar ratio $r$, which are the main
observables that an inflationary model generates, considering
\mbox{$\phi\lsim\langle\phi\rangle$} so the influence of the inflaton to the
potential density is not disregarded during inflation. Similarly, the influence
of $\phi_{\rm end}$ to the value of $N(\phi)$ is retained. The latter effect is
crucial because, as shown below, when \mbox{$\lambda\ll 1$} we have
\mbox{$N\rightarrow N+1/4\sqrt\lambda$}, which can be become significantly
larger than 60.

The quartic hilltop inflation model is:
\begin{equation}
V(\phi)=V_0\left[1-\lambda\left(\frac{\phi}{m_P}\right)^4\right]+\cdots\,,
\label{V}
\end{equation}
where the ellipsis denotes higher order terms which stabilise the potential.
They become important only when the inflaton reaches its VEV; not during
inflation, so we disregard them from now on (but see later).

From the above, we have
\begin{eqnarray}
  V' & = & -4\lambda\frac{V_0}{m_P}\left(\frac{\phi}{m_P}\right)^3
  \label{V'}\\
        {\rm and}\quad V'' & = &
        -12\lambda\frac{V_0}{m_P^2}\left(\frac{\phi}{m_P}\right)^2.
\label{V''}
\end{eqnarray}
Using these, we find the slow-roll parameters
\begin{eqnarray}
\varepsilon=\frac12 m_P^2\left(\frac{V'}{V}\right)^2 & = & 
\frac{8\lambda^2\left(\frac{\phi}{m_P}\right)^6}%
     {\left[1-\lambda\left(\frac{\phi}{m_P}\right)^4\right]^2}\label{eps}\\
     {\rm and}\quad \eta=m_P^2\frac{V''}{V} & = &
     -\,\frac{12\lambda\left(\frac{\phi}{m_P}\right)^2}%
     {1-\lambda\left(\frac{\phi}{m_P}\right)^4}<0\,.
     \label{eta}
\end{eqnarray}

Connecting the inflaton value with the remaining e-folds of inflation, we find
\begin{eqnarray}
\hspace{-.5cm}
  N=\frac{1}{m_P^2}\int_{\phi_{\rm end}}^{\phi(N)}\frac{Vd\phi}{V'} &
  \Rightarrow & \bar N\equiv N+N_{\rm end}
  =\frac{1}{8\lambda}\left[\left(\frac{\phi}{m_P}\right)^{-2}+
    \lambda\left(\frac{\phi}{m_P}\right)^2\right]\label{Nbar}\\
 & \Rightarrow & \left(\frac{\phi}{m_P}\right)^2=4\bar N[Z]\,,
  \label{phiN}
\end{eqnarray}
where 
\begin{equation}
  [Z]\equiv 1
  -\sqrt{1-\frac{1}{Z}}>0\,,
\label{[Z]}
\end{equation}
with
\begin{equation}
Z\equiv 16\lambda\bar N^2>0\,.
\label{Z}
\end{equation}
In the above,
we have defined
\begin{equation}
  N_{\rm end}\equiv
  \frac{1}{8\lambda}\left[\left(\frac{\phi_{\rm end}}{m_P}\right)^{-2}+
    \lambda\left(\frac{\phi_{\rm end}}{m_P}\right)^2\right]\,.
\label{Nend}
\end{equation}  

Let us estimate $N_{\rm end}$. There are two possibilities, which have to do with
which slow-roll parameter is responsible for the termination of inflation:


\bigskip

\noindent
{\bf\boldmath Case A: $|\eta(\phi_{\rm end})|=1$}

\medskip

From Eq.~(\ref{eta}), we find
\begin{equation}
  \left(\frac{\phi_{\rm end}}{m_P}\right)^4+
  12\left(\frac{\phi_{\rm end}}{m_P}\right)^2-\frac{1}{\lambda}=0\,,
\end{equation}
which has only one acceptable solution
\begin{equation}  
  \left(\frac{\phi_{\rm end}}{m_P}\right)^2=
  6\left(\sqrt{1+\frac{1}{36\lambda}}-1\right)\,.
\label{phiendeta}
\end{equation}
Using this in Eq.~(\ref{Nend}), we obtain
\begin{equation}
  N_{\rm end}=\frac{1}{8\lambda}\left[
    \frac{1}{6}\left(\sqrt{1+\frac{1}{36\lambda}}-1\right)^{-1}+
    6\lambda\left(\sqrt{1+\frac{1}{36\lambda}}-1\right)\right]\,.
  \label{Nendetaexact}
\end{equation}  
The above exact result is reduced to the limits
\begin{equation}
  N_{\rm end}\simeq\left\{\begin{array}{cl}
  \frac{1}{4\sqrt\lambda} & {\rm when}\;\lambda\ll 1\\
  & \\
\frac32 & {\rm when}\;\lambda\gg 1
  \end{array}\right.\,.
\label{Nendeta}
\end{equation}


\bigskip

\noindent
{\bf\boldmath Case B: $\varepsilon(\phi_{\rm end})=1$}

\medskip

From Eq.~(\ref{eps}), we find
\begin{equation}
\left[\left(\frac{\phi_{\rm end}}{m_P}\right)+2\sqrt 2\right]
\left(\frac{\phi_{\rm end}}{m_P}\right)^3=\frac{1}{\lambda}\,.
\label{phiendeps}
\end{equation}
This is impossible to solve analytically. However, we can still consider the
limiting values of $\lambda$.
Indeed, if \mbox{$\lambda\ll 1$} then we have
\mbox{$\phi_{\rm end}\gg m_P$} and the above equation gives
\begin{equation}
\left(\frac{\phi_{\rm end}}{m_P}\right)^2=\frac{1}{\sqrt\lambda}\,.
\end{equation}
Inserting this in Eq.~(\ref{Nend}) we find \mbox{$N_{\rm end}=1/4\sqrt\lambda$}.
In the opposite limit \mbox{$\lambda\gg 1$} we have
\mbox{$\phi_{\rm end}\ll m_P$} and Eq.~(\ref{phiendeps}) becomes
\begin{equation}
\left(\frac{\phi_{\rm end}}{m_P}\right)^3=\frac{1}{2\sqrt 2\lambda}\,.
\end{equation}
Inserting this in Eq.~(\ref{Nend}) we obtain
\mbox{$N_{\rm end}=1/4\lambda^{1/3}<1$}.

Thus, in general we have
\begin{equation}
  N_{\rm end}\simeq\left\{\begin{array}{cl}
  \frac{1}{4\sqrt\lambda} & {\rm when}\;\lambda\ll 1\\
  & \\
\frac{1}{4\lambda^{1/3}}<1 & {\rm when}\;\lambda\gg 1
  \end{array}\right.\,.
\label{Nendeps}
\end{equation}

In overall, no matter which slow-roll parameter is taken to end inflation, we
have found that \mbox{$N_{\rm end}\lsim 1$} when \mbox{$\lambda\gg 1$} and
\begin{equation}
N_{\rm end}=\frac{1}{4\sqrt\lambda}\quad{\rm when}\;0<\lambda\ll 1\,.
\label{Nendfin}
\end{equation}
Thus, we see that when \mbox{$\lambda\gg 1$} the contribution of $N_{\rm end}$
to $\bar N$ is negligible so that \mbox{$\bar N\simeq N$}. As we show below,
this is unacceptable so the only possibility is \mbox{$\lambda\ll 1$} for which
\begin{equation}
\bar N=N+\frac{1}{4\sqrt\lambda}\,.
\label{barN}
\end{equation}

To find the observational requirements on the value of $\lambda$ we need to
calculate the spectral index of the generated primordial curvature perturbation.
For the spectral index we have
\begin{equation}
n_s=1-6\varepsilon+2\eta=1-2|\eta|\left(1+3\frac{\varepsilon}{|\eta|}\right)\,.
\label{ns0}
\end{equation}
From Eqs.~(\ref{eta}) and (\ref{phiN}), we find
\begin{equation}
2|\eta|=\frac{3}{\bar N}\frac{Z[Z]}{1-Z[Z]}\,.
\label{2eta}  
\end{equation}
Similarly, using Eqs.~(\ref{eps}) and (\ref{phiN}) we obtain
\begin{equation}
3\frac{\varepsilon}{|\eta|}=
     \frac{2Z[Z]-1}{1-Z[Z]}
     \;\Rightarrow\;
     1+3\frac{\varepsilon}{|\eta|}=\frac{Z[Z]}{1-Z[Z]}
     \,,
     \label{SRratio}
\end{equation}
where we used that
\begin{equation}
1-\lambda\left(\frac{\phi}{m_P}\right)^4=2(1-Z[Z])\,.
\label{brackets}
\end{equation}
Combining Eqs.~(\ref{ns0}), (\ref{2eta}) and (\ref{SRratio}), we find
\begin{equation}
n_s=1-\frac{3}{\bar N}\left(\frac{Z[Z]}{1-Z[Z]}\right)^2\,.
\label{ns}
\end{equation}
The above reduces to \mbox{$n_s=1-3/\bar N$} in the limit \mbox{$Z\gg 1$}.
Disregarding also $N_{\rm end}$ we obtain Eq.~(\ref{nsnaive}). This reveals
the influence of considering \mbox{$\phi\lsim\langle\phi\rangle$}
(corresponding to \mbox{$Z\gsim 1$}) compared to 
\mbox{$\phi\ll\langle\phi\rangle$} (corresponding to \mbox{$Z\gg 1$}), which
was assumed in Ref.~\cite{lythriotto}.

Recasting Eq.~(\ref{ns}), we find
\begin{equation}
\frac{\bar N}{3}(1-n_s)=\left(\frac{Z[Z]}{1-Z[Z]}\right)^2=
\left(\frac{1}{\sqrt{1-\frac{1}{Z}}}\right)^2=\frac{Z}{Z-1}
\label{Zstuff}
\end{equation}
Thus, in view of Eq.~(\ref{Z}), we obtain
\begin{equation}
  16\lambda\bar N^2=Z=\frac{\bar N(1-n_s)}{\bar N(1-n_s)-3}\,.
\label{ZN}
\end{equation}
Note that the above requires that \mbox{$\bar N(1-n_s)>3$} so that \mbox{$Z>0$}.
If \mbox{$\lambda\gg 1$} then \mbox{$\bar N\simeq N$} and the requirement
becomes \mbox{$n_s\leq 1-3/N$}. For \mbox{$N\leq 60$} we find
\mbox{$n_s\leq 0.95$}, which is observationally excluded. Therefore, we must
have \mbox{$\lambda\ll 1$}. In this case $\bar N$ is given by Eq.~(\ref{barN}).

\begin{figure} [h]
\vspace{-9cm}

\centering
\includegraphics[width=\linewidth]{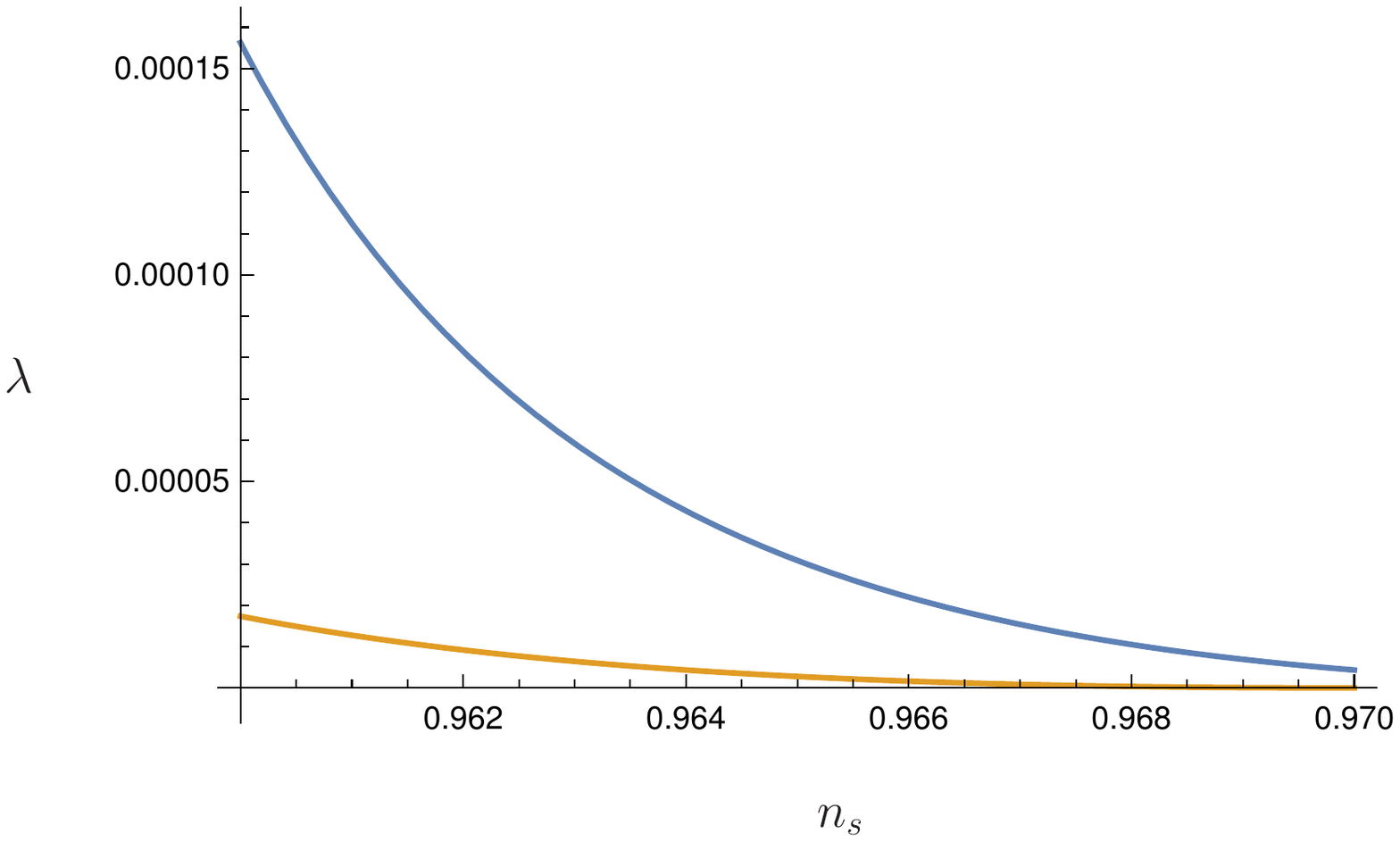} 
\vspace{-5cm}
\caption{The value of $\lambda$ as a function of $n_s$ in the range of interest
  \mbox{$n_s\in[0.960,0.970]$} for different values of $N$ according to
  Eq.~(\ref{sqrtlambda}): \mbox{$N=60$} in the upper curve (blue) and
  \mbox{$N=50$} in the lower curve (orange). It can be seen that,
  when \mbox{$N=50-60$} we have \mbox{$\lambda\lsim 10^{-4}$} in the range of
  interest for $n_s$. This means that the VEV
  \mbox{$\langle\phi\rangle\sim m_P/\lambda^{1/4}\gsim 10\,m_P$} is mildly
  super-Planckian.}  
\label{l4hill}
\end{figure}

Using Eqs.~(\ref{Z}), (\ref{barN}) and (\ref{ZN}), we find
\begin{equation}
  \sqrt\lambda=\frac{2(1-n_s)N-3}{4N[3-(1-n_s)N]} 
\;\Rightarrow\;
\mu\equiv m_P/\lambda^{1/4}=
\left\{\frac{4N[3-(1-n_s)N]}{2(1-n_s)N-3}\right\}^{1/2}m_P\,,
\label{sqrtlambda}
\end{equation}
where $\mu\sim\langle\phi\rangle$ is the inflaton VEV.
This is valid only when \mbox{$\frac32<N(1-n_s)<3$}, because
\mbox{$\sqrt\lambda>0$}. Considering the range \mbox{$N=50-60$} the observations
for the spectral index (\mbox{$n_s=0.965\pm 0.004$} for negligible tensors
\cite{Planck2018}) suggest that \mbox{$\lambda\lsim 10^{-4}$}. This means that
the VEV of the inflaton is super-Planckian
\mbox{$\langle\phi\rangle\sim m_P/\lambda^{1/4}\gsim 10\,m_P$}, which undermines
the perturbative origin of the potential. Values of $\lambda(n_s)$ for a given
choice of $N$ are shown in Fig.~\ref{l4hill}.

Let us discuss the tensor to scalar ratio now. Using Eqs.~(\ref{eps}),
(\ref{phiN}) and (\ref{brackets}), we can write
\begin{equation}
  r=16\varepsilon=\frac{128\lambda^2(\frac{\phi}{m_P})^6}
  {\left[1-\lambda(\frac{\phi}{m_P})^4\right]^2}=
\frac{128\lambda^2\left(4\bar N[Z]\right)^3}{\left[2(1-Z[Z])\right]^2}\,.
\label{r0}
\end{equation}
Combining this with Eq.~(\ref{Zstuff}), we have
\begin{equation}
  r=\frac{32\lambda^2(4\bar N)^3[Z]}{(1-\frac{1}{Z})Z^2}=
\frac{8}{\bar N}\frac{Z[Z]}{Z-1}=\frac83(1-n_s)[Z]\,,
\label{r1}
\end{equation}
where we also used Eq.~(\ref{Z}). Employing the definition of $[Z]$ in
Eq.~(\ref{[Z]})  and Eq.~(\ref{Zstuff}) again, we end up with
\begin{equation}
r=\frac83(1-n_s)\left[1-\frac{\sqrt 3}{\sqrt{(1-n_s)\bar N}}\right]\,.
\label{r}
\end{equation}
Because $r>0$ and the spectrum is red, we expect the expression in the square
brackets in the above to be positive. Using Eq.~(\ref{barN}) and after a little
algebra, this requirement becomes the bound
\mbox{$\sqrt\lambda<\frac14\left(\frac{3}{1-n_s}-N\right)^{-1}$}. In view of
Eq.~(\ref{sqrtlambda}), this bound amounts to \mbox{$n_s>1-\frac{9}{2N}$},
which is satisfied for \mbox{$N=50-60$} and the observed values of $n_s$.

\begin{figure} [h]
\vspace{-3cm}

\centering
\includegraphics[width=\linewidth]{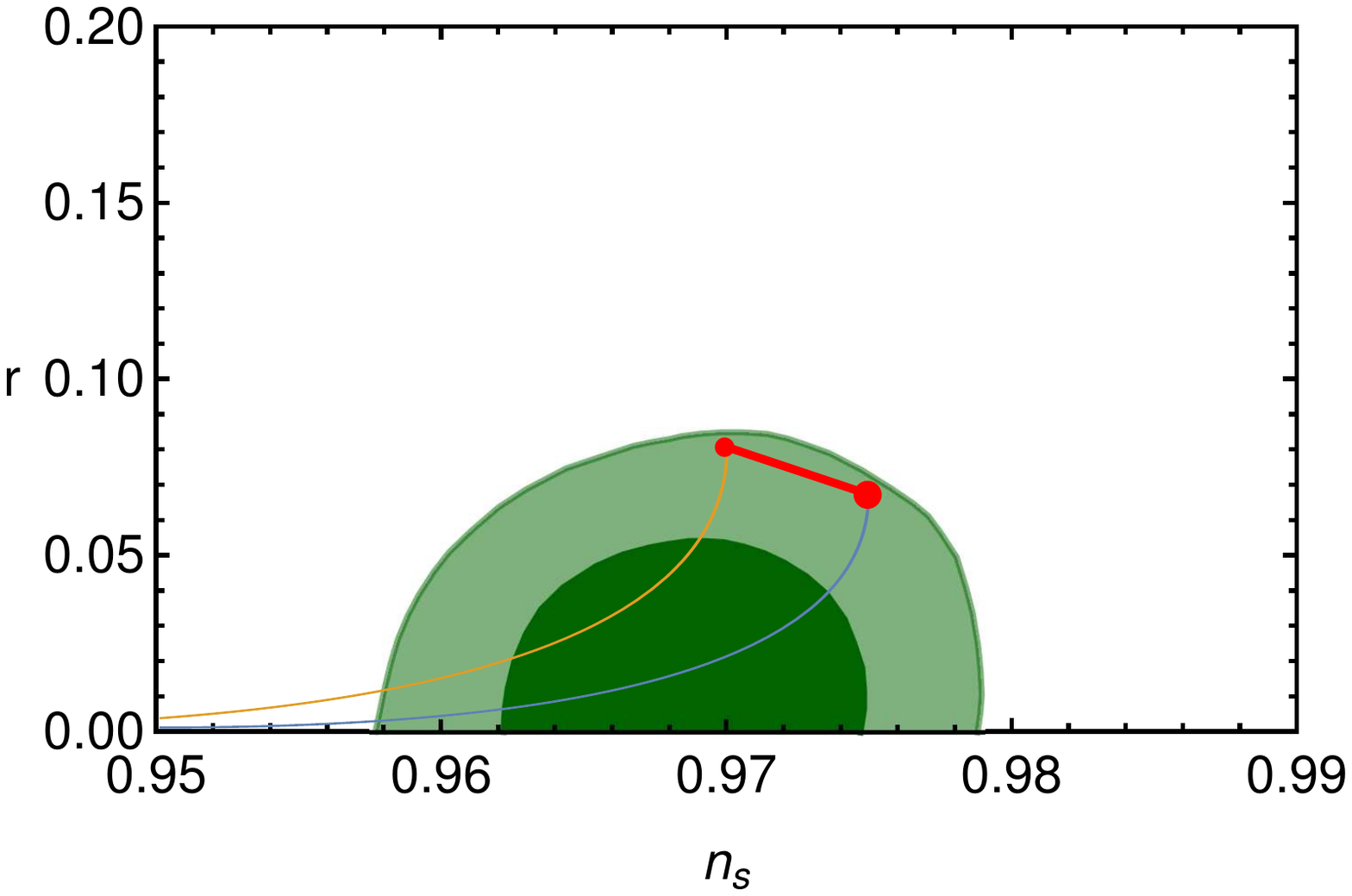} 
\vspace{-11cm}

\caption{The predictions of quartic hilltop inflation for selected values of
  $N$ according to Eq.~(\ref{rfin}): \mbox{$N=60$} in the lower curve (blue) and
  \mbox{$N=50$} in the upper curve (orange), contrasted with the observational
  data of Planck 2015. Values of \mbox{$N\in(50,60)$} correspond to the band
  between the above curves. The band is capped at the slanted solid line (red)
  which depicts the predictions of linear inflation.}  
\label{nsr4hill}
\end{figure}

Using Eqs.~(\ref{Nbar}) and (\ref{sqrtlambda}), we can recast Eq.~(\ref{r}) as
\begin{equation}
  r=\frac83(1-n_s)\left\{1-\frac{\sqrt{3[2(1-n_s)N-3]}}{(1-n_s)N}\right\}\,.
\label{rfin}
\end{equation}
For typical values \mbox{$N=60$} and \mbox{$n_s=0.965$} the above suggests
\mbox{$r=0.0446$}, which will be observable in the near future. Contrasting the
above with observations reproduces the analysis of the Planck Collaboration,
as can be seen by comparing Fig.~\ref{Planck2018} with Fig.~\ref{nsr4hill}.

It is interesting that the limiting values of $r(n_s)$ correspond to the
predictions of linear inflation, with \mbox{$V(\phi)\propto\phi$}. Indeed,
linear inflation suggests \mbox{$n_s=1-\frac{3}{2N}$}. Inserting this value in
Eq.~(\ref{rfin}) one obtains \mbox{$r=\frac{4}{N}$}, which is the prediction of
linear inflation. This has been noticed for some time but it is nice to confirm
it analytically here. In fact, in Ref.~\cite{inflationaris} it is shown that
this limit is attained by all hilltop models of the form in Eq.~(\ref{V0}) with
any \mbox{$q>0$}. The issue was recently explored in Ref.~\cite{linde}, who
argued that hilltop potentials approximate linearity when
\mbox{$V\rightarrow 0$} and the potential is about to become negative unless
stabilised by higher order terms implied by the ellipsis in Eqs.~(\ref{V0}) and
(\ref{V}). But these terms would deform the potential near the VEV, rendering
the form of the potential different from linear.
Therefore, the authors of Ref.~\cite{linde} argue that the approach to
the linear inflation predictions in Fig.~\ref{nsr4hill} is not to be trusted.
How far from this are the predictions of the model trustworthy depends on the
stabilising terms of the potential.

Finally, let us enforce the COBE constraint to estimate $V_0$ in the potential,
such that the model generates the correct magnitude of the curvature
perturbation. In slow-roll inflation we have
\begin{equation}
  \sqrt{\calp_\zeta}=\frac{1}{2\sqrt 3\pi}\frac{V^{3/2}}{m_P^3|V'|}=
  \frac{1}{8\sqrt 3\pi\lambda}\frac{\sqrt{V_0}}{m_P^2}
  \frac{\left[1-\lambda(\frac{\phi}{m_P})^4\right]^{3/2}}{(\frac{\phi}{m_P})^3}=
  \frac{\sqrt 2}{32\sqrt 3\pi\lambda}\frac{\sqrt{V_0}}{m_P^2}
  \left(\frac{1-Z[Z]}{\bar N[Z]}\right)^{3/2}\,,
\label{calp1}
\end{equation}
where we used Eqs.~(\ref{V}) and (\ref{V'}) and then
Eqs.~(\ref{phiN}) and (\ref{brackets}). In view of Eqs.~(\ref{[Z]}), (\ref{Z})
and (\ref{Zstuff}), we can write
\begin{equation}
\frac{1-Z[Z]}{\bar N[Z]}=16\lambda\left(\frac{3\bar N}{1-n_s}\right)^{1/2}\,.
\label{Zfin}
\end{equation}
Combining the above with Eq.~(\ref{calp1}) we find
\begin{equation}
\frac{V_0}{m_P^4}=\frac{3\pi^2\calp_\zeta}{8\lambda}
\left(\frac{1-n_s}{3\bar N}\right)^{3/2}\,.
\label{calp}
\end{equation}
Taking the values \mbox{$\calp_\zeta=2.1\times 10^{-9}$} and
\mbox{$n_s\simeq 0.965$} with \mbox{$\lambda\simeq 10^{-4}$} and \mbox{$N=60$}
(so \mbox{$\bar N\simeq 85$}, cf. Eq.~(\ref{Nbar})) we find
\mbox{$V_0^{1/4}\sim 10^{16}\,$GeV}, which is the
GUT-scale as expected.

This concludes our analytical study of quartic hilltop inflation. Our main
results are the expressions in Eqs.~(\ref{sqrtlambda}) and (\ref{rfin}). The
latter is shown to reproduce exactly the numerical results, as can be seen in
Fig.~\ref{nsr4hill}. We found that, for quartic hilltop inflation to work we
need a mildly super-Planckian VEV of order
\mbox{$\langle\phi\rangle\gsim 10\,m_P$}. A sub-Planckian VEV results in the
traditional expression in Eq.~(\ref{nsnaive}), which is ruled out as it
gives rise to \mbox{$n_s\lsim 0.95$}. The performance of the model, when the VEV
is super-Planckian, is crucially determined by the
contribution of the value $\phi_{\rm end}$ of the inflaton when inflation is
terminated, which can add significantly to the number of e-folds which
correspond to the cosmological scales, as demonstrated by Eq.~(\ref{Nbar}).
It has to be noted that the predictions of the model are not trustworthy when
approaching the results of linear inflation (see Fig.~\ref{nsr4hill}) because
the effect of the stabilising terms in the potential cannot be ignored then.
However, this limit is already disfavoured by observations
(see Fig.~\ref{Planck2018}).

After this work came out, it was realised that some analytic treatment of
quartic hilltop inflation is also done in Ref.~\cite{chiamin}. The author
obtains a parametric expression of $r(n_s)$, which is arguably not as
straightforward as Eq.~(\ref{rfin}).

I would like to thank David Sloan for discussions. KD was
supported (in part) by the Lancaster-Manchester-Sheffield Consortium for
Fundamental Physics under STFC grant: ST/L000520/1.

\end{document}